\documentclass[twocolumn]{aastex631}
\usepackage{epsfig}
\usepackage{rotating}
\usepackage[T1]{fontenc}
\usepackage{ae,aecompl}
\usepackage{tikz}

\usepackage{amsmath}
\usepackage{amssymb}
\usepackage{graphicx}
\usepackage{color}
\usepackage{natbib}

\usepackage{commath}
\usepackage{hyperref}

\newcommand{\Ms}{{\ensuremath{M_{\odot} }}}
\newcommand{\Zs}{{\ensuremath{Z_{\odot} }}}

\submitjournal{AAS journals}

\shorttitle{LRDs are DCBH galaxies}
\shortauthors{Whalen et al.}

\begin{document}

\title{Little Red Dots are Direct-Collapse Black Hole-Forming Galaxies}

\correspondingauthor{Daniel J. Whalen}
\email{dwhalen1999@gmail.com}

\author[0000-0001-6646-2337]{Daniel J. Whalen}
\affiliation{Institute of Cosmology and Gravitation, Portsmouth University, Dennis Sciama Building, Portsmouth PO1 3FX}

\correspondingauthor{Muhammad A. Latif}
\email{latifne@gmail.com}

\author[0000-0003-2480-0988]{Muhammad A. Latif}
\affiliation{Physics Department, College of Science, United Arab Emirates University, PO Box 15551, Al-Ain, UAE}

\author[0009-0009-0009-1364]{Konstantinos Topalakis}
\affiliation{Department of Physics, Aristotle University of Thessaloniki, Thessaloniki 54124, Greece}

\author{Fergus Cullen}
\affiliation{Institute for Astronomy, University of Edinburgh, Royal Observatory, Blackford Hill, Edinburgh EH9 3HJ, UK}

\author{Devesh Nandal}
\affiliation{Center for Astrophysics, Harvard and Smithsonian, 60 Garden St, Cambridge, MA 02138, USA}

\author{Sadegh Khochfar}
\affiliation{Institute for Astronomy, University of Edinburgh, Royal Observatory, Blackford Hill, Edinburgh EH9 3HJ, UK}

\begin{abstract}

"Little Red Dots" (LRDs) at $4 < z < 8$ are one of the most challenging discoveries by {\em JWST} to date because their distinctive V-shaped spectra and compact morphologies defy conventional astrophysical interpretation.  Previous attempts to explain LRDs as compact stellar systems, heavily-cocooned black holes with differential flows, supermassive stars, or more exotic objects like 'black-hole stars' either cannot show how they formed, explain the origin of the dense shells needed for the absorption features in their spectra, or account for their observed abundances or inferred lifetimes. Here we show that LRDs are simply direct-collapse black hole galaxies in which the BH is still shrouded by the massive disk that created it.  Our cosmological simulations yield spectra that are good matches to those of LRDs because high densities at the center of the disk trap X-rays from the BH and produce the observed Balmer absorption features while allowing UV, optical and reprocessed IR flux to partly escape.  The host galaxy forms a dense 10$^8$ \Ms\ cluster of stars with a radius of 150 pc next to the BH, consistent with observations of LRDs.  Our models reproduce a wide variety of LRD spectra from typical objects like RUBIES-EGS-42046 at $z = 5.28$ to those with the strongest Balmer breaks such as MoM-BH$^*$-1 at $z = 7.76$ and those at the highest redshifts like CAPERS-LRD-z9 at $z = 9.29$.  

\end{abstract}


\section{Introduction}

Over 300 "Little Red Dots" (LRDs) have now been discovered at redshifts $z \sim$ 4 - 8 \citep{lrd1,lrd2,lrd4,lrd3}.  Their distinctive V-shaped spectra, compact morphologies ($\sim$ 100 - 200 pc), and high luminosities with little or no X-ray emission challenge conventional astrophysical explanations.  A growing body of work now suggests that the red continua of LRDs are not simply due to dust-reddened starlight.  Their spectra often turn over near the Balmer limit, producing the characteristic V-shaped spectral energy distribution (SED) in which the rest-UV remains relatively blue while the rest-optical continuum is very red \citep{Setton2024}.  These features can arise if the central source is hidden inside an optically thick, dust-poor reprocessing layer whose emergent photosphere is cool, with $T_{\rm eff} \sim 4000$--$6000$ K \citep{Liu2025}.  In this picture, the observed color is set by the opacity structure of dense hydrogen-rich gas rather than by an old stellar population.  Such a photosphere could naturally mask a much hotter ionizing source and convert its radiation into the red optical continuum seen by the {\em James Webb Space Telescope} ({\em JWST}).

Line spectra provide independent evidence for dense gas very close to the power source.  Several LRDs show broad Balmer emission, extreme Balmer decrements, Balmer absorption, and in some cases P-Cygni-like structure or blueshifted absorption components \citep{Inayoshi2025,Torralba2025,Matthee2026}.  These features are difficult to produce in low-density nebulae because the $n=2$ level of hydrogen must be collisionally populated for strong Balmer absorption to appear.  The inferred densities are therefore very high, typically $n \gtrsim 10^9$--$10^{11}\ {\rm cm}^{-3}$, and the required columns are often near or above the Compton-thick regime.  The blueshifted absorption components are especially important because, in the usual P-Cygni interpretation, absorption blueward of the systemic Balmer or helium line traces dense gas with an outward line of sight velocity component.

This has led to a physical picture in which many LRDs are compact engines embedded in dense cocoons or shells of gas.  In the highest-quality spectra, the broad Balmer and helium wings may be shaped by electron scattering rather than by purely virial motions in a classical broad-line region \citep{Rusakov2026}.  The same cocoon can also explain why LRDs are luminous in the rest-frame optical but weak in X-rays, radio emission, and hot dust emission compared with ordinary unobscured AGN \citep{latif24b,Maiolino2025,Setton2025, Rusakov2026}.  The observational constraints therefore point to a compact source surrounded by a high covering-factor reprocessing medium, with a cool photosphere outside and a dense ionized or partially ionized layer inside.

Supermassive primordial stars \citep[SMSs;][]{nan26a} with prior mass ejections that build up dense shells  \citep{nan26b} can recreate many of the spectral features of LRDs but their short lifetimes (1 - 2 Myr) cannot account for the numbers of LRDs observed to date.  Hypothetical 'black-hole stars' invoked to account for more extreme examples of LRDs \citep{nai26} require the existence of quasi-stars \citep[e.g.,][]{begel08} that have never been found to form in any numerical simulation of supermassive stellar evolution to date \citep{hos13,tyr17,hle18b,herr23a,nan25a} because the onset of triple-$\alpha$ and CNO burning in the contracting protostar prevents the early collapse of the core into a BH.  Massive, compact stellar clusters \citep{bag24} cannot easily account for the extremely high gas densities needed for the Balmer self-absorption features in LRD spectra or how the clusters themselves formed.  

Direct-collapse black holes (DCBHs) due to the collapse of SMSs are a more promising route to the rise of LRDs because they are persistent sources that last for tens of Myr and are more likely to appear in relatively narrow {\em JWST} fields of view \citep[e.g.,][]{bl03,wf12,bar18,sur18a,sur19a,smidt18,wet20b,latif21a,latif22b,jeon26}.  Their observed numbers to date are also on par with the number density of DCBHs found in cosmological simulations at similar eras \citep{hab16} and recent observations of broad-line active galactic nuclei (AGNs) in high-$z$ galaxies \citep{fei26}.  But their successes so far in reproducing the V-shaped spectrum of LRDs either do not account for the compact stellar clusters inferred from their SEDs or depend on simplified 1D geometries that cannot rule out the possibility of X-ray breakout at early times, which would seriously limit the number of LRDs observed today \citep{rus26,snep26,pac26}.  Nevertheless, DCBHs can solve many problems associated with LRDs because they form in extremely high densities at the center of the massive accretion disk that created the SMS, which have been previously shown to prevent X-ray breakout in cosmological simulations \citep[see Figure 1 of][]{wet20b} and their host halos exhibit centrally-peaked density profiles that are conducive to the formation of compact clusters of stars at later times.  Recent cosmological simulations also show that DCBH host galaxies exhibit the high BH / stellar masses observed in LRDs \citep{jeon25,latif26a,chon26}.

We present cosmological simulations with radiation hydrodynamics that show that DCBH formation in primordial halos can produce LRD spectra and their compact stellar populations.  In Section 2 we describe our numerical simulations of DCBH birth and rise of compact stellar populations in the host galaxy at later times.  We describe our spectrum calculations and compare our SEDs to those of several LRDs in Section 3 and conclude in Section 4.


\begin{figure*} 
\plottwo{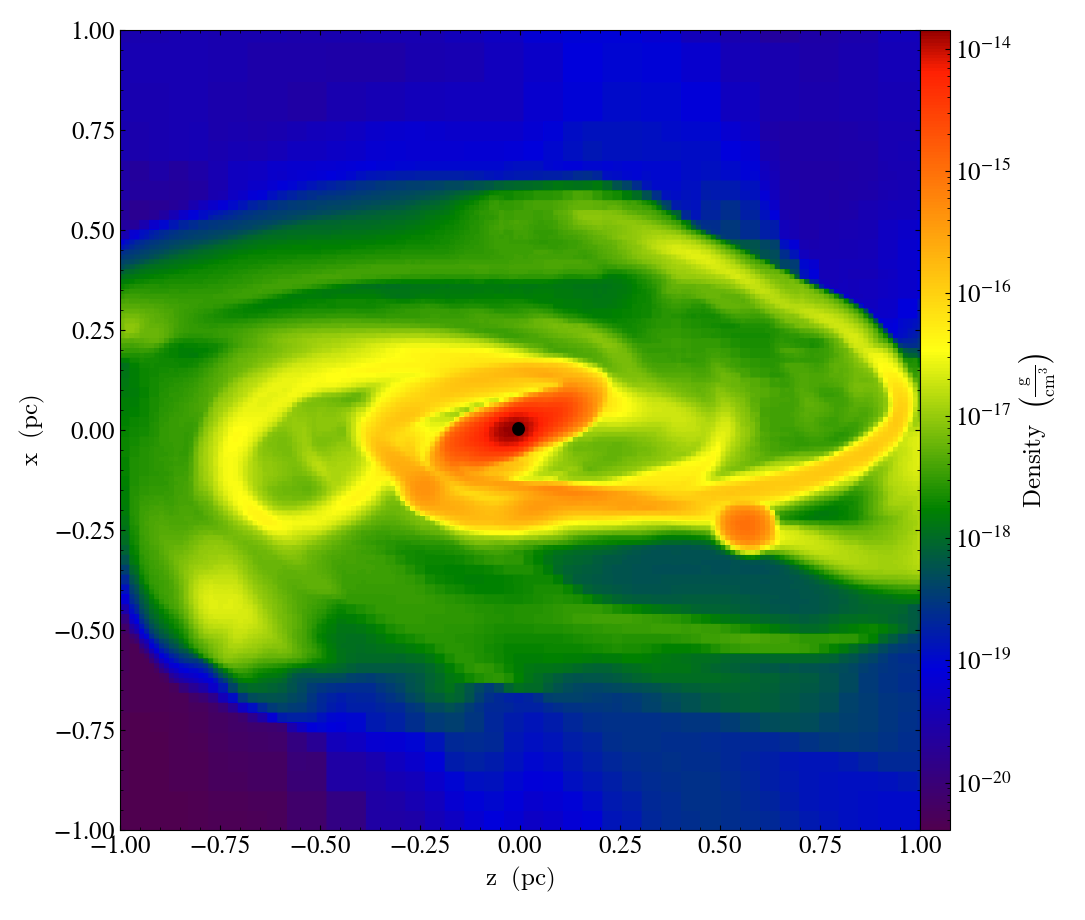}{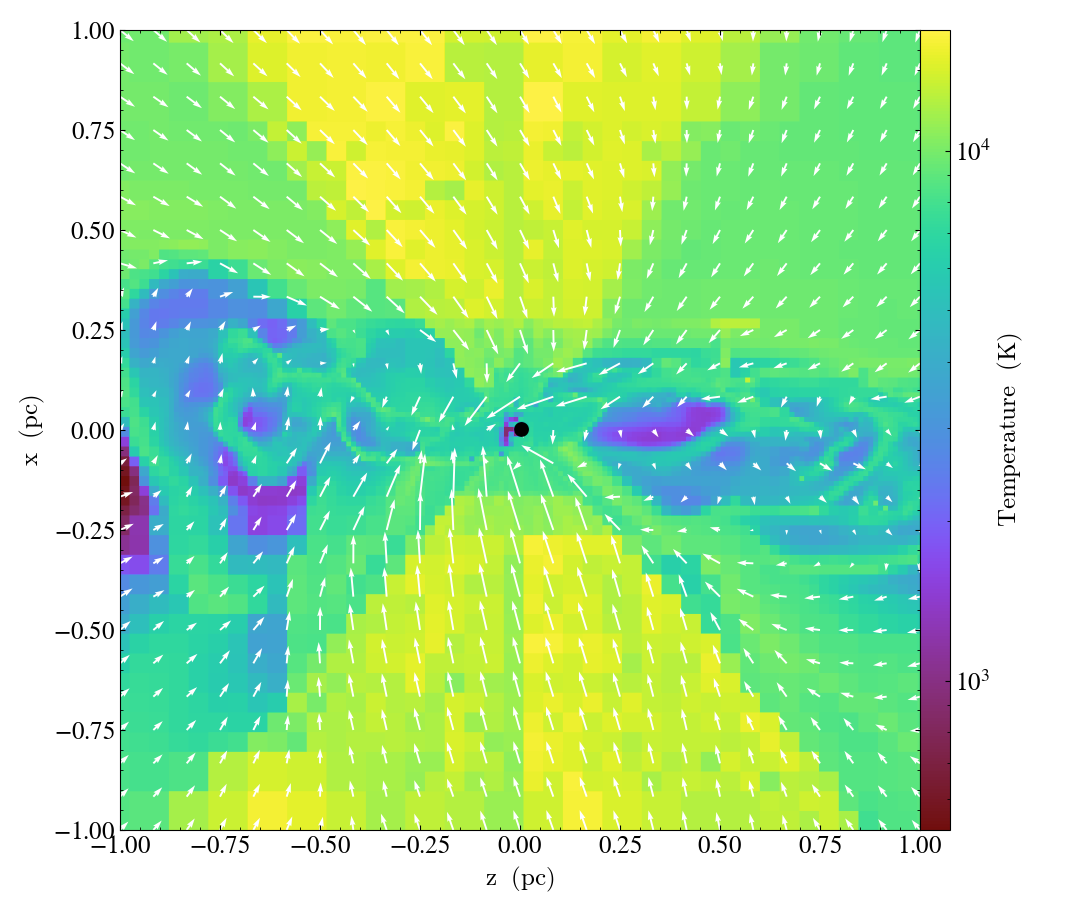}
\caption{The DCBH and its accretion disk.  Left:  densities, which reach 10$^{10}$ cm$^{-3}$ in the vicinity of the BH shown as the black dot.  Right:  temperatures and velocity vectors in the disk, in which the H II region is trapped at the center on scales below those of the image.}
\vspace{0.1in}
\label{fig:dcbh} 
\end{figure*}

\section{Numerical Methods} \label{sec:method}

No single simulation can resolve X-rays from a DCBH deep in its accretion disk while evolving its host galaxy over the times required to follow the rise of stellar populations, so we compute LRD spectra in two stages.  First, we initialize and evolve a 10$^5$ \Ms\ DCBH at the center of an atomically-cooled accretion disk like those in which SMSs form but with a luminosity of $\sim$ 10$^{45}$ erg s$^{-1}$, corresponding to a 10$^7$ \Ms\ BH accreting at the Eddington limit, to determine if the disk can trap X-rays at the BH masses inferred for LRDs.  This procedure yields the lower limit for X-ray breakout because the disk and host halo in reality will have grown to higher masses by the time the BH reaches LRD masses \citep[see, e.g.,][]{pat23a}.  We then use the densities, temperatures and electron mass fractions of the ultracompact \ion{H}{2} regions deep in the disk to compute DCBH spectra with Cloudy.  We note that previous studies have shown that Ly-$\alpha$ photons can also be trapped at similar densities and temperatures during Pop III star formation \citep{latif11} and DCBH formation \citep{aaron17}, creating pressures that can also affect the evolution of the cloud.

Next, we extract the ages, masses and metallicities of the stars in the DCBH host galaxy simulated by \citet{latif26a} when the BH has reached $\sim$ 10$^7$ \Ms\ to compute the stellar component of the LRD spectrum with the Binary Population and Stellar Synthesis (BPASS) code.  This latter simulation was the first to evolve a DCBH for several hundred Myr while fully resolving Pop II and III star formation in the host galaxy with X-rays from the BH and ionizing UV, winds and SNe due to the stars but not the accretion disk itself.  The DCBH and stellar spectra from our fiducial host galaxy are then merged to produce LRD SEDs.  This approach enables us to self-consistently model both components of LRD spectra without being held to the shortest timescales dictated by processes deep in the accretion disk.  

\subsection{DCBH / Disk Simulation}

We evolve the DCBH in its accretion disk with Enzo, an adaptive mesh refinement (AMR) cosmology code with gas and dark matter dynamics, ionizing photon transport, and nonequilibrium primordial gas chemistry and cooling \citep{enzo}.  Enzo has a third-order piecewise-parabolic method for hydrodynamics \citep{bryan95} with an HLLC Riemann solver for better stability with strong shocks and rarefaction waves \citep{toro94}.  Dark matter is evolved with an $N$-body particle-mesh method \citep{efs85,couch91} and self-gravity is calculated with a multigrid Poisson solver.  We use six-species nonequilibrium primordial gas chemistry (H, He, e$^-$, H$^+$, He$^+$, He$^{2+}$) with cooling due to collisional excitations and ionizations of H and He, recombinations of H and He, bremsstrahlung emission, and inverse Compton cooling by the cosmic microwave background (CMB).   

We propagate X-rays through the simulation volume with the MORAY raytracing radiation transport code \citep{moray}.  Energy from electrons due to X-ray photoionizations is partitioned between thermalization in the gas and secondary ionizations \citep{Shull85,kim11} and radiation pressure due to photoionizations is included in updates to the momentum equations.  We assume monoenergetic 2 keV BH photons, consistent with the average quasar spectral energy distribution \citep{sos04} and observations that show 90\% of the X-ray flux from BHs at $z >$ 6 to be at 0.5 - 2 keV \citep{nan17,smidt18}.  This choice produces the lowest mass at which X-rays from the DCBH can break out of the disk and terminate the LRD spectrum because 2 keV photons have the longest mean free path through the gas over this range in energy.   


\begin{figure} 
\plotone{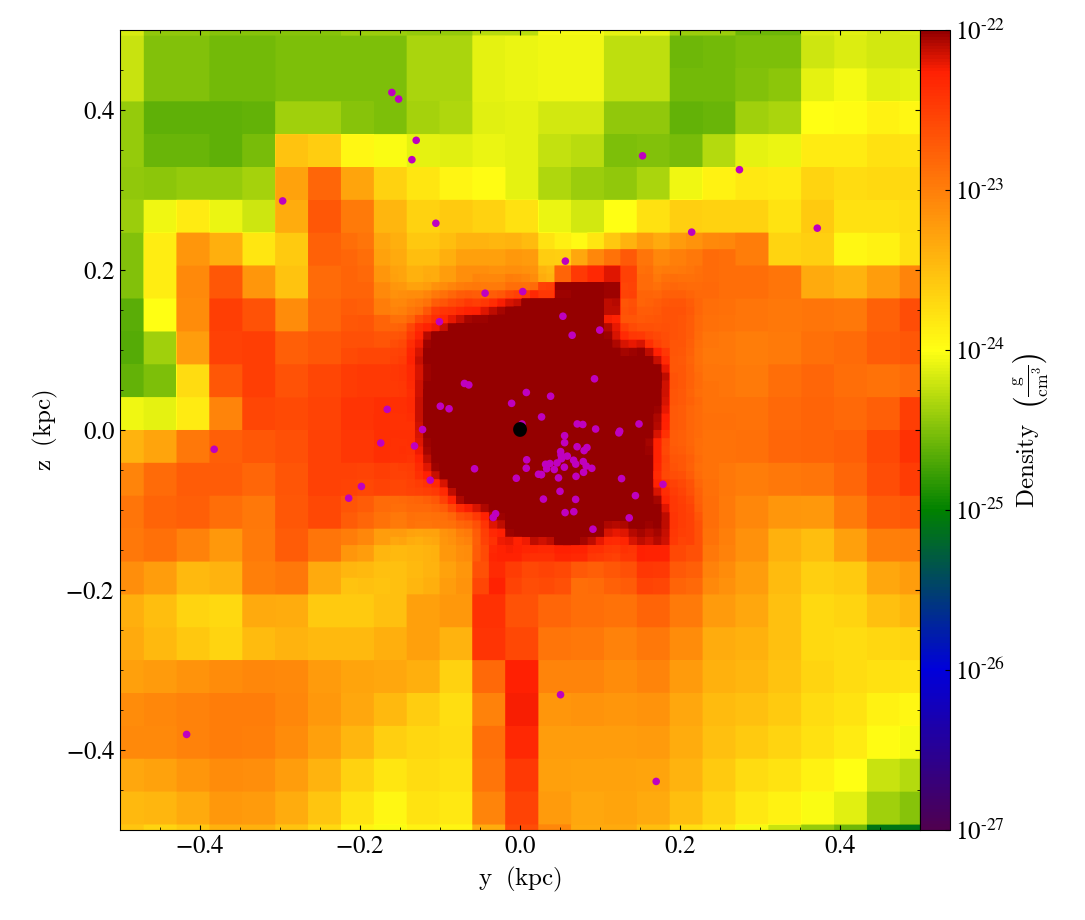}
\caption{The compact massive stellar cluster enclosing a DCBH in its host halo at $z =$ 12.7 from the Enzo simulation in \citet{latif26a}.  The stellar mass is $1.2 \times 10^8$ \Ms, the great majority of which is due to chemically-enriched Pop II stars.  Purple dots mark the positions of small Pop II star particles, which are mostly clustered within a 150 pc radius, and the black spot shows the position of the DCBH.}
\vspace{0.1in}
\label{fig:spop} 
\end{figure}

The BH is treated as a sink particle and point source of X-ray photons, and we assume Eddington-limited Bondi-Hoyle accretion \citep{latif20b}:
\begin{equation}
\dot{M}_{\rm BH}  =  min \left( \frac{4 \pi G^2  M_{\rm BH}^2 \rho_{\rm B}}{(c_{\rm s}+v_{\rm rel})^3},  \frac{4 \pi G M_{\rm BH} m_{\rm p}}{\epsilon_{\rm r} \sigma_{\rm T} c} \right)
\end{equation}
where $\dot{M}_{\rm BH}$ is the accretion rate, $\epsilon_{\rm r} = 0.1$ is the radiative efficiency, $G$ is the gravitational constant, $M_{\rm BH}$ the black hole mass, $\rho_{\rm B}$ is the density at the Bondi radius, $c_{\rm s}$ is the sound speed, $v_{\rm rel}$ is the gas velocity relative to the BH, $m_{\rm p}$ is the proton mass and $\sigma_{\rm T}$ is the Thomson scattering cross-section. The Bondi accretion radius is
\begin{equation}
R_{\rm B} = \frac{2 G M_{\rm BH}}{c_{\rm s}^2},
\label{eq3}
\end{equation}
which is 20 - 30 pc for the DCBH and is well resolved by our minimum cell size of 0.009 pc.  The BH luminosity is  
\begin{equation}
L_{\mathrm{bol}} = \epsilon \dot{M}_{\mathrm{BH}} c^2,  
\end{equation}
where the radiative efficiency, $\epsilon =$ 0.1.  We scale this luminosity up by a factor of 40 to fluxes of $\sim$ 10$^{45}$ erg s$^{-1}$, those of a 10$^7$ \Ms\ BH radiating at about the Eddington limit, to determine if the disk still traps X-rays from it after it has grown to this mass.  

The SMS that created the DCBH considered here did not alter flows in its vicinity prior to collapse. To reach 10$^5$ \Ms, an SMS must accrete at average rates $\gtrsim$ 0.03 \Ms\ yr$^{-1}$ \citep[Figure 4 of][]{tyr17} and stellar evolution calculations show that such stars are cool ($\sim$ 5500 K) and red over their short lifetimes, producing little ionizing UV radiation that can drive outflows that could reduce or halt accretion onto them \citep{hle18b,herr23a,nan24d}.  Our simulations thus capture the true state of flows onto the DCBH at birth.  

Our simulation box is 1 $h^{-1}$ Mpc with a 256$^3$ root grid and three nested grids that span 20\% of the top grid and are centered on the halo for an maximum initial resolution of 1024$^3$.  We initialize the run at $z =$ 150 with Gaussian random perturbations from MUSIC \citep{hahn11} with \textit{Planck} second-year cosmological parameters:  $\Omega_{\mathrm M} = 0.308$, $\Omega_\Lambda = 0.691$, $\Omega_{\mathrm b}h^2 = 0.0223$, $\sigma_8 =$ 0.816, $h = $ 0.677 and $n =$ 0.968 \citep{planck2}.  We allow up to 15 levels of refinement after the onset of collapse for a maximum spatial resolution of 0.009 pc.  Our refinement criteria are: (1) a baryonic overdensity threshold of 4; (2) a dark matter mass refinement condition of 0.0625 $\rho_{\rm DM} r^{\ell \alpha}$, where $\rho_{\rm DM}$ is the dark matter density, $r = 2$ is the refinement factor, $\ell$ is the refinement level, and $\alpha = -0.3$ ensures super-Lagrangian refinement; and (3) enforcing a minimum of 32 cells per Jeans length \citep{latif20b, latif22b}.  The BH is evolved for 0.3 Myr after initialization in the disk.

\subsection{Stellar Population}

We show the stellar population of the DCBH host galaxy at $z =$ 12.7 in Figure~\ref{fig:spop}, when the BH has reached $\sim$ 10$^7$ \Ms\ in the \citet{latif26a} Enzo simulation.  The stars have a total mass of $1.2 \times 10^8$ \Ms, 2000 \Ms\ of which are Pop III stars, and at this point the halo has a metallicity of 0.02 \Zs.  They are confined to a radius of $\sim$ 150 pc at the center of the halo, where gas densities are highest ($\sim$ 10$^4$ cm$^{-3}$), demonstrating that DCBH host galaxies can exhibit the compact morphologies of LRDs.  We extract the masses, ages and metallicities of the stars at this redshift for our BPASS spectra.  The dense stellar cluster is a simple result of the $\sim r^{-2.2}$ dark matter / gas density profile of the halo, which concentrates fuel for star formation at its center as shown in Figure~\ref{fig:spop}.  Consistent with recent observations of lensed LRDs in which the blue stellar component is separated from the red AGN component by 50 - 100 pc \citep[e.g.,][]{bag26,gol26}, the DCBH is offset by the center of the cluster by about 75 pc.

\begin{deluxetable}{lccc}
\tabletypesize{\scriptsize}
\tablecaption{SED Fit Parameters \label{tbl-1}}
\tablehead{
\colhead{LRD)} & \colhead{log $L_{\rm bol}$} & \colhead{$A_{\rm v}$ (BH)} & 
\colhead{$A_{\rm v}$ (stars)}}
\startdata
RUBIES-EGS-42046    & 45.15  & 0.4     & 0.05   \\
RUBIES-UDS-154183  & 45.25  & 0.5     & 0.5     \\
MoM-BH*-1                   & 45.35  & 0.9     & 0.4    \\
RUBIES-EGS-14295    & 45.5    & 1.6     & 0.1    \\
CAPERS-LRD-z9         & 45.25  & 1.05   & 0.2    \\
\enddata
\end{deluxetable}

\section{DCBH Disk Spectra} 

The halo begins to atomically cool when it reaches a mass of 1.44 $\times$ 10$^7$ \Ms\ at $z =$ 17.  It forms a massive disk at its center, at which time we turn on X-rays from the DCBH at the center of the disk.  We show an overhead view of the disk in the left panel of Figure~\ref{fig:dcbh} at 100 kyr, when it is approximately 1 pc in diameter.  As we show in the right panel of Figure~\ref{fig:dcbh}, X-rays from the DCBH are confined to the center of the disk.  Although temperatures in the ultracompact (UC) \ion{H}{2} region reach $\sim$ 10$^7$ K because of X-ray heating, it remains trapped deep in the halo throughout the run because of gas densities that exceed 10$^{10}$ cm$^{-3}$ and large ram pressures due to heavy infall that can reach 1 \Ms\ yr$^{-1}$, as shown by the velocity vectors.  There are fluctuations due to turbulent flows in the disk but bolometric luminosities average $\sim$ 0.85 $L_\mathrm{Edd}$.  


\begin{figure} 
\begin{center}
\begin{tabular}{cc}
\epsfig{file=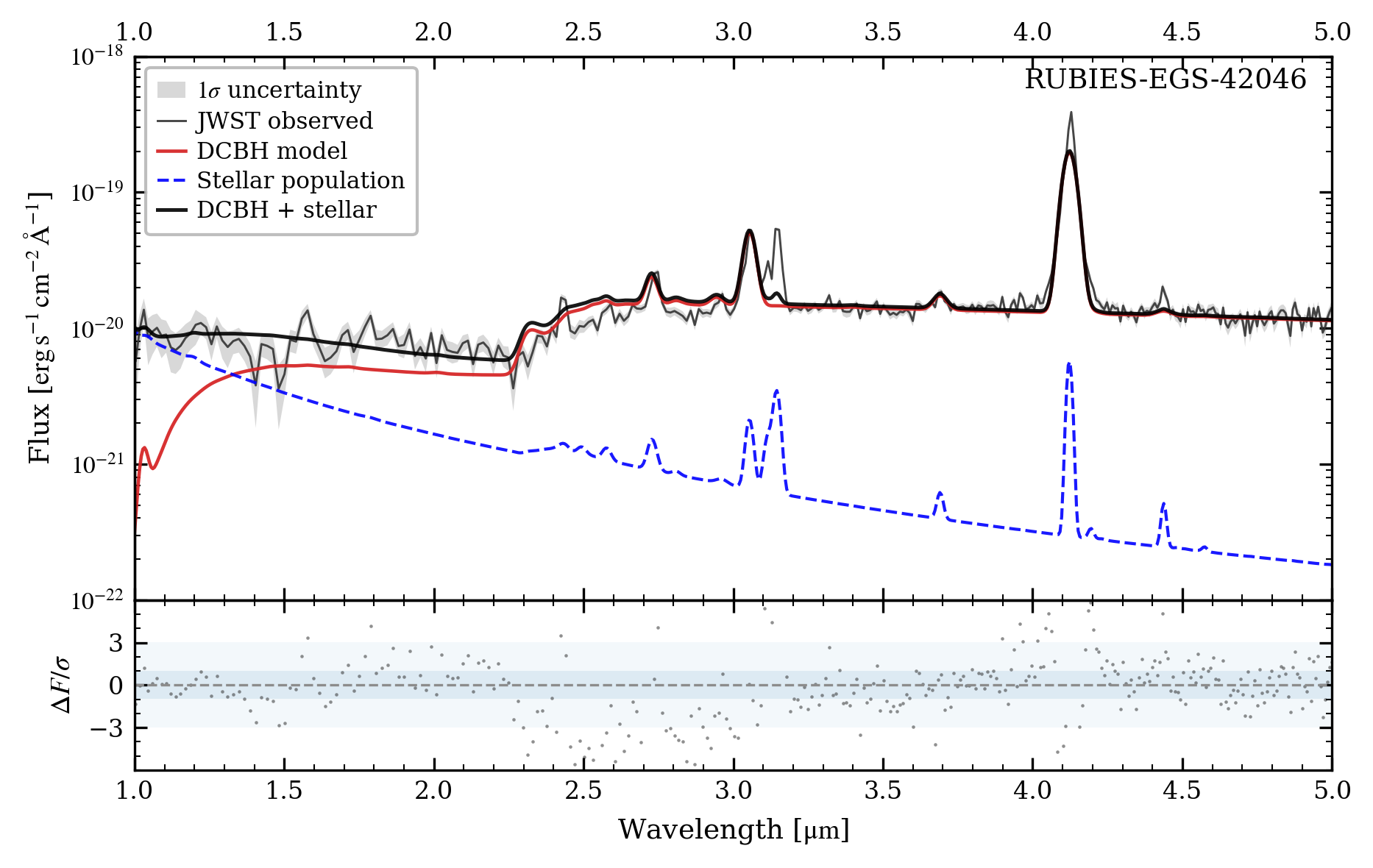,width=0.95\linewidth,clip=}   \\
\epsfig{file=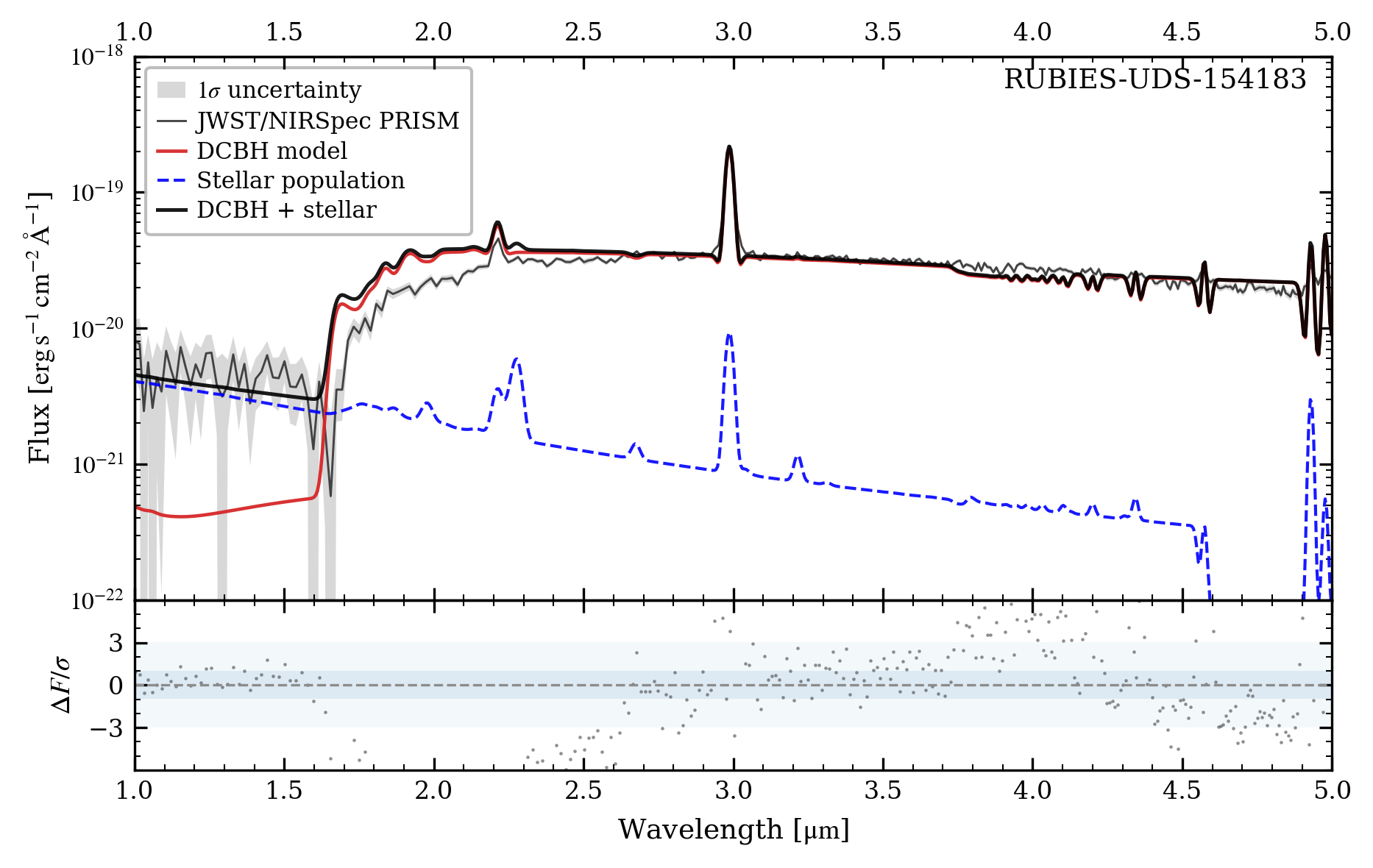,width=0.95\linewidth,clip=}   \\
\epsfig{file=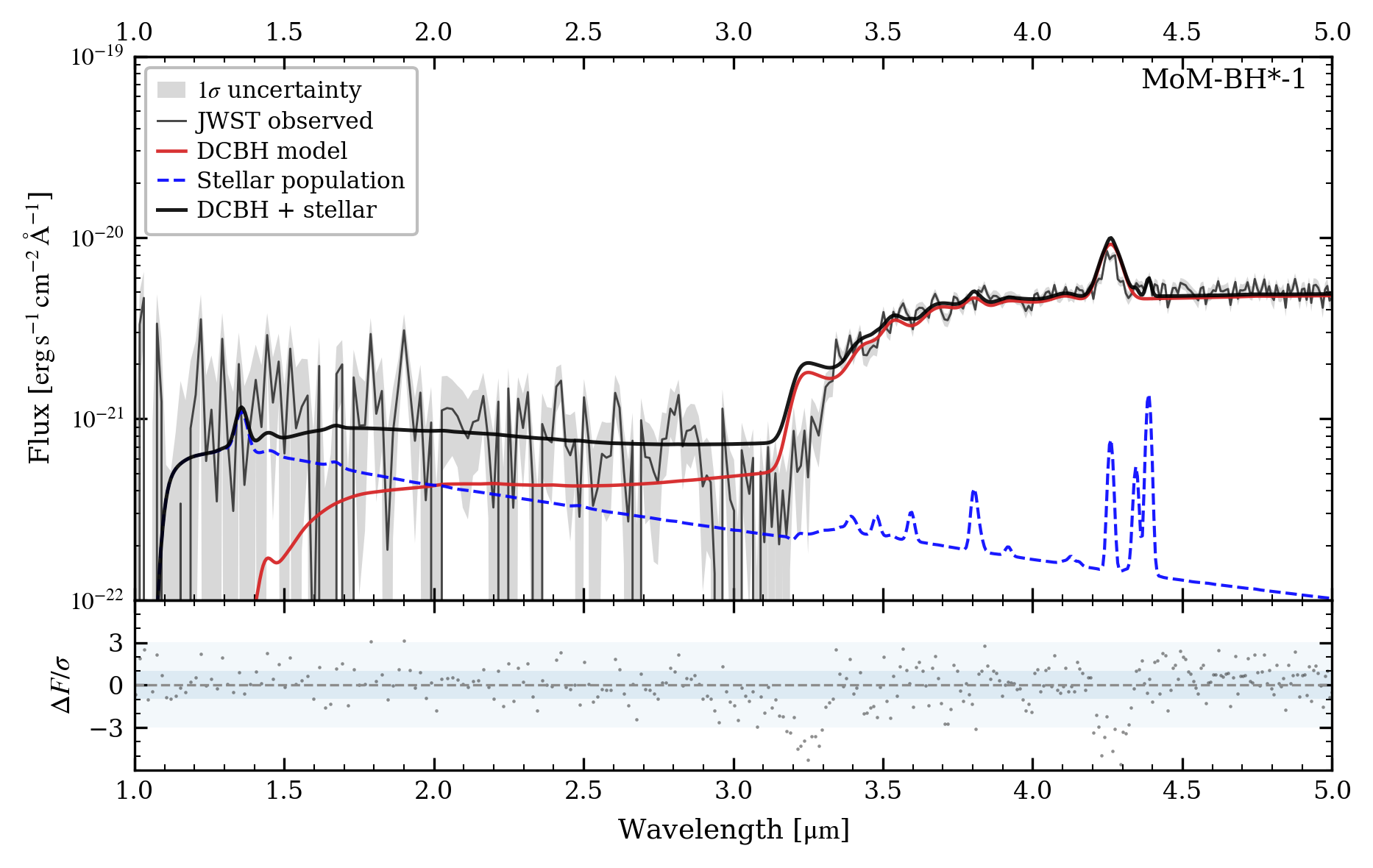,width=0.95\linewidth,clip=}   \\
\epsfig{file=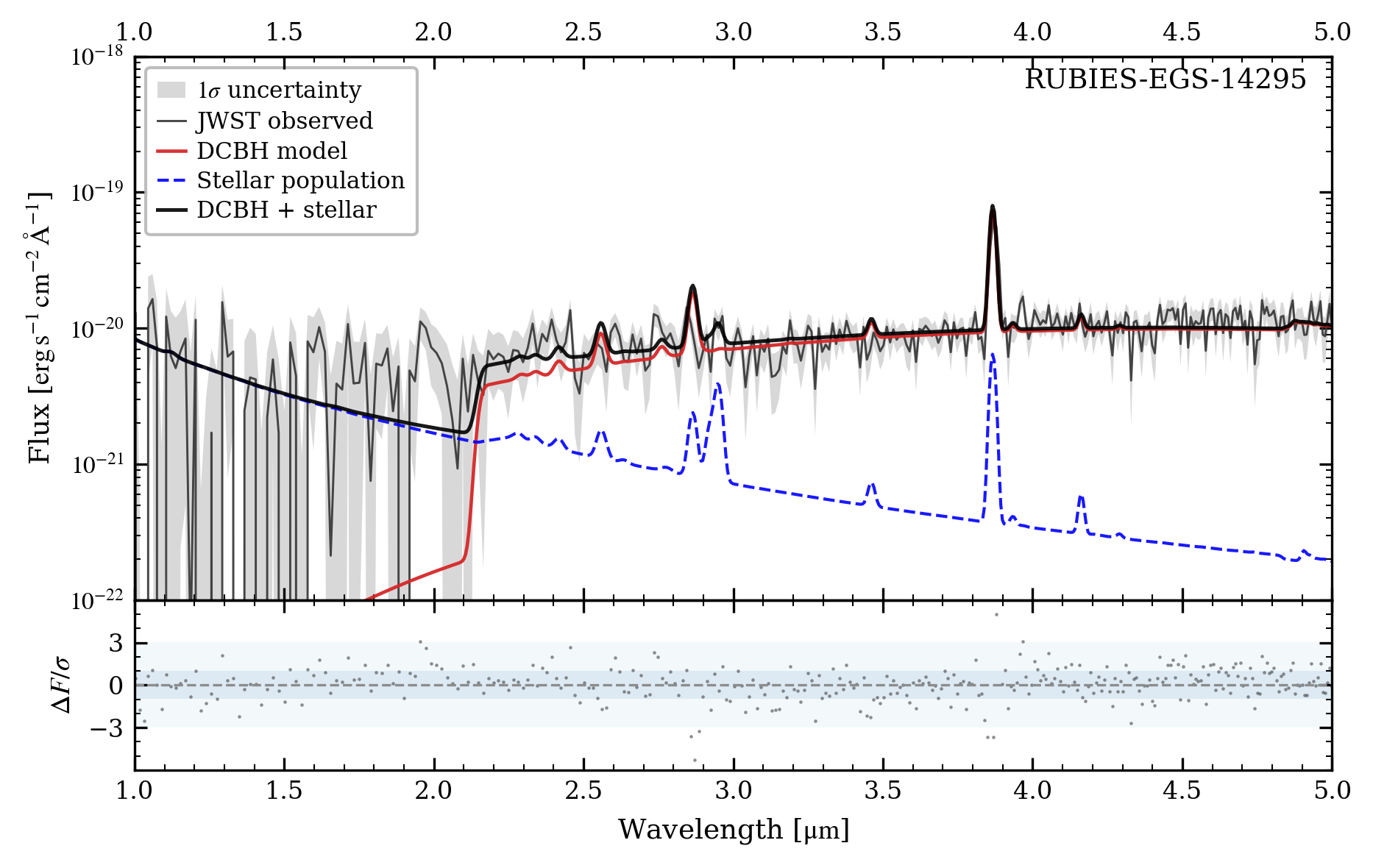,width=0.95\linewidth,clip=}  
\end{tabular}
\end{center}
\caption{Our spectral fits to a variety of LRD SEDs.  From top to bottom: RUBIES-EGS-42046 at $z =$ 5.28, RUBIES-UDS-154183 at $z =$ 3.55, MoM-BH*-1 at $z =$ 7.76, RUBIES-EGS-14295 at $z =$ 4.89.}
\vspace{0.1in}
\label{fig:spec} 
\end{figure}

We calculate spectra with Cloudy v23.01 \citep{Guna23} from the properties of the environment of the DCBH in our Enzo model and the ionizing DCBH spectrum from \citet{L25} normalized to specific bolometric luminosities.  We computed a grid of models over bolometric luminosities log $L_{\rm bol, BH} =$ 44.5 - 46.0, inner cloud radii log $R =$ 16.8 - 18.5 cm, hydrogen number densities log $n_{\rm H} =$ 10 - 11.5 cm$^{-3}$, turbulent velocities log $v_{\rm turb} =$ 1.5 - 3.5 km s$^{-1}$, and column densities log $N_{\rm H} =$ 23.5 - 26.0.  Spherical geometry is assumed with a fixed gas metallicity log $Z/\Zs\ =$ -2.0.  While our Enzo simulations show that free electron mass fractions due to the UC \ion{H}{2} region of the DCBH exist in its vicinity, Cloudy does not perform line broadening due to scattering by free electrons.  We therefore convolve our DCBH spectra with a Gaussian kernel to account for broad-line emission, followed by convolution with the wavelength-dependent observed line-spread function and resampling onto the instrumental wavelength grid.  Other physical processes like turbulent gas velocities can produce similar broadening.  We calculated SEDs for the stellar population with BPASS v.2.3 \citep{Eld17,Byrne22} from the masses, ages and metallicities of the individual star particles in the DCBH host halo in \citet{latif26a} at $z =$ 12.7, again assuming a spherical geometry but with a fixed gas density of $n_{\mathrm{H}} = 10^3 \, \mathrm{cm^{-3}}$ and a metallicity log $Z/\Zs\ =$ -2.0.  The stellar and DCBH spectra were summed to produce the final output spectrum.  The DCBH and stellar dust extinction coefficients and $L_{\rm bol}$ used in the fits are shown in Table~\ref{tbl-1}.  

\begin{figure} 
\plotone{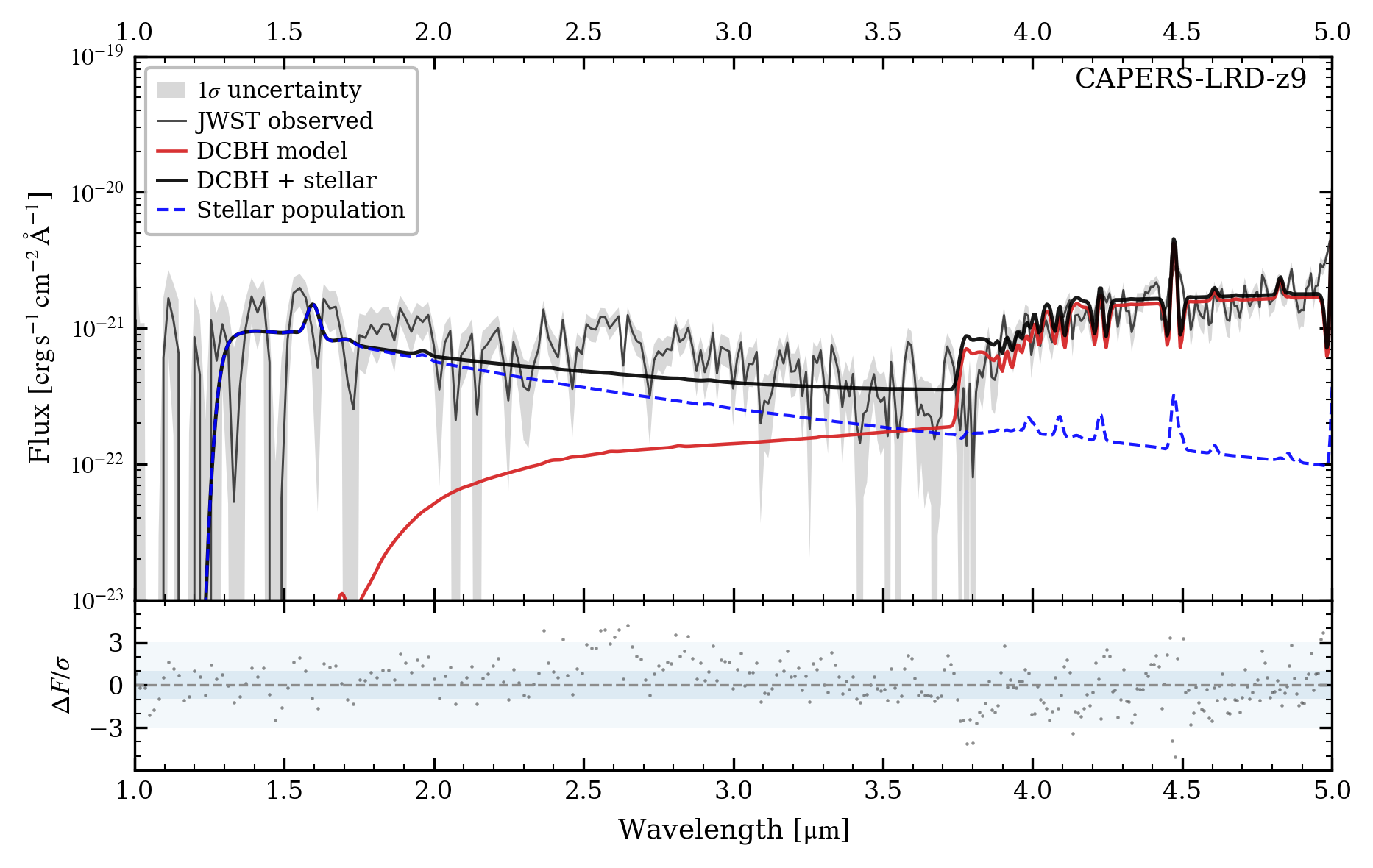}
\caption{Our spectral fit to CAPERS-LRD-z9 at $z =$ 9.29.}
\vspace{0.1in}
\label{fig:capers} 
\end{figure}

We show fits of our spectra to five fiducial LRD SEDs in Figures~\ref{fig:spec} and \ref{fig:capers}:  RUBIES-EGS-42046, a fairly typical LRD at $z =$ 5.28 \citep{hvid25,pac26}, RUBIES-UDS-154183, an LRD with an exceptional, cliff-like jump at $z =$ 3.55 \citep{grf25,snep26}, MoM-BH*-1, an outlier with the strongest observed Balmer break at $z =$ 7.76 \citep{nai26}, RUBIES-EGS-14295, an extended source with a v-shaped spectrum at $z =$ 4.89 \citep{hvid25}, and CAPERS-LRD-z9, the highest redshift LRD found to date at $z =$ 9.29 in which the BH / stellar mass ratio may be as high as 0.3 \citep{tyl25}.  The DCBH can account for nearly all the RUBIES-EGS-42046 spectrum down to $\sim$ 1.5 $\mu$m because stars only contribute a few percent of the flux above this wavelength but they are necessary below 1.5 $\mu$m because the DCBH spectrum falls off so sharply there.  \citet{pac26} are able to fit this spectrum with just a DCBH and some dust extinction due to a small ($\sim$ 6000 \Ms) population of stars but preprocess their AGN spectra prior to use in Cloudy so it differs from ours in spite of their common origin \citep{yue13}.

The other four LRDs all require a stellar component to match the blue components of their SEDs below 1.5 - 3.5 $\mu$m.  In particular, we find that the bolometric flux of a 10$^7$ \Ms\ BH radiating at the Eddington limit and a stellar mass of $1.2 \times 10^8$ \Ms\ yield a good fit to the CAPERS-LRD-z9 spectrum, whose corresponding masses had previously been inferred to be $\sim$ $3 \times 10^8$ \Ms\ and 10$^9$ \Ms, respectively \citep{tyl25}.  Such masses would be difficult to explain from cosmological structure formation because SF would somehow have to have been suppressed as the BH rose to large masses.  Our $M_{\rm BH} / M_*$ ratio is much more in line with those inferred for other LRDs and overmassive BH galaxies such as UHZ1 \citep{cet23,Bod23,Gould23} and GHZ9 \citep{Atek23,Cast23,Kov24}.

\section{Conclusion}


We find that DCBH host galaxies can account for a wide range of LRD spectra, from typical ones like RUBIES-EGS-42046 at $z =$ 5.28 to outliers with strong Balmer breaks such as MoM-BH*-1 at z = 7.76 and even those with extreme BH / stellar mass ratios such as CAPERS-LRD-z9 at $z = $ 9.29, without resorting to exotic or hypothetical objects such as quasi-stars or black-hole stars.  They can also create the massive, compact stellar populations required for LRD morphologies and that are thought to contribute the bluer component of most LRD spectra.  The clusters are a natural result of star formation in $r^{-2.2}$ dark matter density profiles in which fuel for star formation is concentrated at the center of the halo.  The recent discovery that the red and blue components in gravitationally-lensed LRDs are separated by 50 - 100 pc \citep[e.g.,][]{bag26,gol26} reinforces the notion that they are DCBH galaxies, as the BH and stellar cluster in our simulation exhibit similar offsets \citep[see, e.g.,][]{regan17a}.

Although the DCBH and stars in our models formed at $z \sim$ 17 and 13, too early to be any of the LRDs found so far, cosmological simulations indicate that DCBH formation may have peaked at lower redshifts, $z \sim$ 8 - 10 \citep{hab16}, from which they then could have evolved into LRDs by $4 < z < 8$.  DCBH disk and host galaxy morphologies (and thus spectra) at these lower redshifts would have been similar to those at the higher redshifts in our models and could explain the observational characteristics of LRDs found to date, especially given that their luminosity functions are consistent with the number densities of DCBHs at $z \sim$ 8 - 10.  In \citet{hab16}, DCBH production peaks over this redshift range and then falls off as the cosmos becomes enriched by metals.  This may explain why the number of LRDs sharply drops below $z \sim$ 4 because as the DCBHs born at $z \sim$ 8 - 10 grow in mass over time, their X-rays finally break out of their natal disks and the LRD spectra disappear, without new DCBHs to replace them.  Furthermore, the failure of X-rays to break out of the disk even at effective BH masses of $\sim$ 10$^7$ \Ms\ suggests that LRDs are objects that can persist for tens of Myr, their inferred lifetimes from observations \citep[e.g.,][]{sun26}, given Salpeter times for BHs growing at the Eddington limit.  

Given the numbers of LRDs found to date, it is clear that {\em JWST} could find them at much earlier stages because can detect 10$^5$ \Ms\ DCBHs at up to $z \sim$ 20 \citep{bar18,wet20b}, whose X-rays would certainly be trapped by the accretion disk.   Although dimmer, they would retain the Balmer breaks and absorption features of LRDs but have less prominent blue components because of lower total stellar masses in the host galaxy.  Although the radio fluxes predicted for LRDs are generally too low for detection by current telescopes \citep[consistent with current observations that LRDs are radio quiet;][]{wet21a,latif25a} they could be found by the Square Kilometer Array (SKA) and Next Generation Very Large Array (ngVLA).  Radio observations may therefore also probe the formation pathways of LRDs in the coming decade.

\begin{acknowledgments}

We thank the anonymous referee, whose comments improved the quality of this paper.  MAL thanks the UAEU for funding via UPAR grant No. G00005454. FC acknowledges support from a UKRI Frontier Research Guarantee Grant (PI Cullen; grant reference EP/X021025/1). DN was supported by the Swiss National Science Fund (SNSF) Postdoctoral Fellowship, grant number: P500-2235464.  SK acknowledges funding via STFC Small Grant ST/Y001133/1.
                     
\end{acknowledgments}

\bibliography{refs.bib}

\begin{thebibliography}{}
\expandafter\ifx\csname natexlab\endcsname\relax\def\natexlab#1{#1}\fi
\providecommand{\url}[1]{\href{#1}{#1}}
\providecommand{\dodoi}[1]{doi:~\href{http://doi.org/#1}{\nolinkurl{#1}}}
\providecommand{\doeprint}[1]{\href{http://ascl.net/#1}{\nolinkurl{http://ascl.net/#1}}}
\providecommand{\doarXiv}[1]{\href{https://arxiv.org/abs/#1}{\nolinkurl{https://arxiv.org/abs/#1}}}

\bibitem[{{Atek} {et~al.}(2023){Atek}, {Chemerynska}, {Wang}, {Furtak},
  {Weibel}, {Oesch}, {Weaver}, {Labb{\'e}}, {Bezanson}, {van Dokkum}, {Zitrin},
  {Dayal}, {Williams}, {Nannayakkara}, {Price}, {Brammer}, {Goulding}, {Leja},
  {Marchesini}, {Nelson}, {Pan}, \& {Whitaker}}]{Atek23}
{Atek}, H., {Chemerynska}, I., {Wang}, B., {et~al.} 2023, \mnras, 524, 5486,
  \dodoi{10.1093/mnras/stad1998}

\bibitem[{{Baggen} {et~al.}(2026){Baggen}, {van Dokkum}, {Labb{\'e}}, \&
  {Brammer}}]{bag26}
{Baggen}, J. F.~W., {van Dokkum}, P., {Labb{\'e}}, I., \& {Brammer}, G. 2026,
  \apjl, 1006, L14, \dodoi{10.3847/2041-8213/ae82fe}

\bibitem[{{Baggen} {et~al.}(2024){Baggen}, {van Dokkum}, {Brammer}, {de
  Graaff}, {Franx}, {Greene}, {Labb{\'e}}, {Leja}, {Maseda}, {Nelson}, {Rix},
  {Wang}, \& {Weibel}}]{bag24}
{Baggen}, J. F.~W., {van Dokkum}, P., {Brammer}, G., {et~al.} 2024, \apjl, 977,
  L13, \dodoi{10.3847/2041-8213/ad90b8}

\bibitem[{{Barrow} {et~al.}(2018){Barrow}, {Aykutalp}, \& {Wise}}]{bar18}
{Barrow}, K. S.~S., {Aykutalp}, A., \& {Wise}, J.~H. 2018, Nature Astronomy, 2,
  987, \dodoi{10.1038/s41550-018-0569-y}

\bibitem[{{Begelman} {et~al.}(2008){Begelman}, {Rossi}, \&
  {Armitage}}]{begel08}
{Begelman}, M.~C., {Rossi}, E.~M., \& {Armitage}, P.~J. 2008, \mnras, 387,
  1649, \dodoi{10.1111/j.1365-2966.2008.13344.x}

\bibitem[{{Bogd{\'a}n} {et~al.}(2023){Bogd{\'a}n}, {Goulding}, {Natarajan},
  {Kov{\'a}cs}, {Tremblay}, {Chadayammuri}, {Volonteri}, {Kraft}, {Forman},
  {Jones}, {Churazov}, \& {Zhuravleva}}]{Bod23}
{Bogd{\'a}n}, {\'A}., {Goulding}, A.~D., {Natarajan}, P., {et~al.} 2023, Nature
  Astronomy, \dodoi{10.1038/s41550-023-02111-9}

\bibitem[{{Bromm} \& {Loeb}(2003)}]{bl03}
{Bromm}, V., \& {Loeb}, A. 2003, \apj, 596, 34, \dodoi{10.1086/377529}

\bibitem[{{Bryan} {et~al.}(1995){Bryan}, {Norman}, {Stone}, {Cen}, \&
  {Ostriker}}]{bryan95}
{Bryan}, G.~L., {Norman}, M.~L., {Stone}, J.~M., {Cen}, R., \& {Ostriker},
  J.~P. 1995, Computer Physics Communications, 89, 149,
  \dodoi{10.1016/0010-4655(94)00191-4}

\bibitem[{{Bryan} {et~al.}(2014){Bryan}, {Norman}, {O'Shea}, {Abel}, {Wise},
  {Turk}, {Reynolds}, {Collins}, {Wang}, {Skillman}, {Smith}, {Harkness},
  {Bordner}, {Kim}, {Kuhlen}, {Xu}, {Goldbaum}, {Hummels}, {Kritsuk}, {Tasker},
  {Skory}, {Simpson}, {Hahn}, {Oishi}, {So}, {Zhao}, {Cen}, {Li}, \& {Enzo
  Collaboration}}]{enzo}
{Bryan}, G.~L., {Norman}, M.~L., {O'Shea}, B.~W., {et~al.} 2014, \apjs, 211,
  19, \dodoi{10.1088/0067-0049/211/2/19}

\bibitem[{{Byrne} {et~al.}(2022){Byrne}, {Stanway}, {Eldridge}, {McSwiney}, \&
  {Townsend}}]{Byrne22}
{Byrne}, C.~M., {Stanway}, E.~R., {Eldridge}, J.~J., {McSwiney}, L., \&
  {Townsend}, O.~T. 2022, \mnras, 512, 5329, \dodoi{10.1093/mnras/stac807}

\bibitem[{{Castellano} {et~al.}(2023{\natexlab{a}}){Castellano}, {Fontana},
  {Treu}, {Merlin}, {Santini}, {Bergamini}, {Grillo}, {Rosati}, {Acebron},
  {Leethochawalit}, {Paris}, {Bonchi}, {Belfiori}, {Calabr{\`o}}, {Correnti},
  {Nonino}, {Polenta}, {Trenti}, {Boyett}, {Brammer}, {Broadhurst}, {Caminha},
  {Chen}, {Filippenko}, {Fortuni}, {Glazebrook}, {Mascia}, {Mason}, {Menci},
  {Meneghetti}, {Mercurio}, {Metha}, {Morishita}, {Nanayakkara}, {Pentericci},
  {Roberts-Borsani}, {Roy}, {Vanzella}, {Vulcani}, {Yang}, \& {Wang}}]{cet23}
{Castellano}, M., {Fontana}, A., {Treu}, T., {et~al.} 2023{\natexlab{a}},
  \apjl, 948, L14, \dodoi{10.3847/2041-8213/accea5}

\bibitem[{{Castellano} {et~al.}(2023{\natexlab{b}}){Castellano}, {Fontana},
  {Treu}, {Merlin}, {Santini}, {Bergamini}, {Grillo}, {Rosati}, {Acebron},
  {Leethochawalit}, {Paris}, {Bonchi}, {Belfiori}, {Calabr{\`o}}, {Correnti},
  {Nonino}, {Polenta}, {Trenti}, {Boyett}, {Brammer}, {Broadhurst}, {Caminha},
  {Chen}, {Filippenko}, {Fortuni}, {Glazebrook}, {Mascia}, {Mason}, {Menci},
  {Meneghetti}, {Mercurio}, {Metha}, {Morishita}, {Nanayakkara}, {Pentericci},
  {Roberts-Borsani}, {Roy}, {Vanzella}, {Vulcani}, {Yang}, \& {Wang}}]{Cast23}
---. 2023{\natexlab{b}}, \apjl, 948, L14, \dodoi{10.3847/2041-8213/accea5}

\bibitem[{{Chon} {et~al.}(2026){Chon}, {Hirano}, {Ishiyama}, {Chang}, \&
  {Springel}}]{chon26}
{Chon}, S., {Hirano}, S., {Ishiyama}, T., {Chang}, S.-J., \& {Springel}, V.
  2026, \nat, 657, 621, \dodoi{10.1038/s41586-026-10985-8}

\bibitem[{{Couchman}(1991)}]{couch91}
{Couchman}, H.~M.~P. 1991, \apjl, 368, L23, \dodoi{10.1086/185939}

\bibitem[{{de Graaff} {et~al.}(2025){de Graaff}, {Rix}, {Naidu}, {Labb{\'e}},
  {Wang}, {Leja}, {Matthee}, {Katz}, {Greene}, {Hviding}, {Baggen}, {Bezanson},
  {Boogaard}, {Brammer}, {Dayal}, {van Dokkum}, {Goulding}, {Hirschmann},
  {Maseda}, {McConachie}, {Miller}, {Nelson}, {Oesch}, {Setton}, {Shivaei},
  {Weibel}, {Whitaker}, \& {Williams}}]{grf25}
{de Graaff}, A., {Rix}, H.-W., {Naidu}, R.~P., {et~al.} 2025, \aap, 701, A168,
  \dodoi{10.1051/0004-6361/202554681}

\bibitem[{{Efstathiou} {et~al.}(1985){Efstathiou}, {Davis}, {White}, \&
  {Frenk}}]{efs85}
{Efstathiou}, G., {Davis}, M., {White}, S.~D.~M., \& {Frenk}, C.~S. 1985,
  \apjs, 57, 241, \dodoi{10.1086/191003}

\bibitem[{{Eldridge} {et~al.}(2017){Eldridge}, {Stanway}, {Xiao}, {McClelland},
  {Taylor}, {Ng}, {Greis}, \& {Bray}}]{Eld17}
{Eldridge}, J.~J., {Stanway}, E.~R., {Xiao}, L., {et~al.} 2017, \pasa, 34,
  e058, \dodoi{10.1017/pasa.2017.51}

\bibitem[{{Fei} {et~al.}(2026){Fei}, {Fujimoto}, {Naidu}, {Chisholm}, {Atek},
  {Brammer}, {Asada}, {Berg}, {Bromm}, {Furtak}, {Greene}, {Hsiao}, {Jeon},
  {Kokorev}, {Matthee}, {Natarajan}, {Pan}, {Richard}, {Saldana-Lopez},
  {Schaerer}, {Volonteri}, \& {Zitrin}}]{fei26}
{Fei}, Q., {Fujimoto}, S., {Naidu}, R.~P., {et~al.} 2026, \apj, 1003, 244,
  \dodoi{10.3847/1538-4357/ae6248}

\bibitem[{{Golubchik} {et~al.}(2026){Golubchik}, {Furtak}, {Allingham},
  {Zitrin}, {Akins}, {Kokorev}, {Fujimoto}, {Abdurro'uf}, {Amor{\'\i}n},
  {Bauer}, {Bezanson}, {Brada{\v{c}}}, {Bradley}, {Brammer}, {Chisholm}, {Coe},
  {Conselice}, {Dayal}, {Dessauges-Zavadsky}, {Diego}, {Faisst}, {Fei},
  {Ferguson}, {Finkelstein}, {Frye}, {Gonz{\'a}lez-Otero}, {Greene},
  {Harikane}, {Hsiao}, {Inayoshi}, {Jim{\'e}nez-Teja}, {Knudsen}, {Koekemoer},
  {Labb{\'e}}, {Lucas}, {Magdis}, {Matthee}, {Messa}, {Naidu}, {Nakane},
  {Noirot}, {Pan}, {Papovich}, {Richard}, {Ricotti}, {Robbins}, {Stark}, {Sun},
  {Treu}, {Tripodi}, {Vanzella}, {Willott}, \& {Windhorst}}]{gol26}
{Golubchik}, M., {Furtak}, L.~J., {Allingham}, J. F.~V., {et~al.} 2026, \aap,
  710, A226, \dodoi{10.1051/0004-6361/202558407}

\bibitem[{{Goulding} {et~al.}(2023){Goulding}, {Greene}, {Setton}, {Labbe},
  {Bezanson}, {Miller}, {Atek}, {Bogd{\'a}n}, {Brammer}, {Chemerynska},
  {Cutler}, {Dayal}, {Fudamoto}, {Fujimoto}, {Furtak}, {Kokorev}, {Khullar},
  {Leja}, {Marchesini}, {Natarajan}, {Nelson}, {Oesch}, {Pan}, {Papovich},
  {Price}, {van Dokkum}, {Wang}, {Weaver}, {Whitaker}, \& {Zitrin}}]{Gould23}
{Goulding}, A.~D., {Greene}, J.~E., {Setton}, D.~J., {et~al.} 2023, \apjl, 955,
  L24, \dodoi{10.3847/2041-8213/acf7c5}

\bibitem[{{Greene} {et~al.}(2024){Greene}, {Labbe}, {Goulding}, {Furtak},
  {Chemerynska}, {Kokorev}, {Dayal}, {Volonteri}, {Williams}, {Wang}, {Setton},
  {Burgasser}, {Bezanson}, {Atek}, {Brammer}, {Cutler}, {Feldmann}, {Fujimoto},
  {Glazebrook}, {de Graaff}, {Khullar}, {Leja}, {Marchesini}, {Maseda},
  {Matthee}, {Miller}, {Naidu}, {Nanayakkara}, {Oesch}, {Pan}, {Papovich},
  {Price}, {van Dokkum}, {Weaver}, {Whitaker}, \& {Zitrin}}]{lrd2}
{Greene}, J.~E., {Labbe}, I., {Goulding}, A.~D., {et~al.} 2024, \apj, 964, 39,
  \dodoi{10.3847/1538-4357/ad1e5f}

\bibitem[{{Greene} {et~al.}(2026){Greene}, {Setton}, {Furtak}, {Naidu},
  {Volonteri}, {Dayal}, {Labbe}, {van Dokkum}, {Bezanson}, {Brammer}, {Cutler},
  {Glazebrook}, {de Graaff}, {Hirschmann}, {Hviding}, {Kokorev}, {Leja}, {Liu},
  {Ma}, {Matthee}, {Nanayakkara}, {Oesch}, {Pan}, {Price}, {Spilker}, {Wang},
  {Weaver}, {Whitaker}, {Williams}, \& {Zitrin}}]{lrd3}
{Greene}, J.~E., {Setton}, D.~J., {Furtak}, L.~J., {et~al.} 2026, \apj, 996,
  129, \dodoi{10.3847/1538-4357/ae1836}

\bibitem[{{Gunasekera} {et~al.}(2023){Gunasekera}, {van Hoof}, {Chatzikos}, \&
  {Ferland}}]{Guna23}
{Gunasekera}, C.~M., {van Hoof}, P. A.~M., {Chatzikos}, M., \& {Ferland}, G.~J.
  2023, Research Notes of the American Astronomical Society, 7, 246,
  \dodoi{10.3847/2515-5172/ad0e75}

\bibitem[{{Habouzit} {et~al.}(2016){Habouzit}, {Volonteri}, {Latif}, {Dubois},
  \& {Peirani}}]{hab16}
{Habouzit}, M., {Volonteri}, M., {Latif}, M., {Dubois}, Y., \& {Peirani}, S.
  2016, \mnras, 463, 529, \dodoi{10.1093/mnras/stw1924}

\bibitem[{{Haemmerl{\'e}} {et~al.}(2018){Haemmerl{\'e}}, {Woods}, {Klessen},
  {Heger}, \& {Whalen}}]{hle18b}
{Haemmerl{\'e}}, L., {Woods}, T.~E., {Klessen}, R.~S., {Heger}, A., \&
  {Whalen}, D.~J. 2018, \mnras, 474, 2757, \dodoi{10.1093/mnras/stx2919}

\bibitem[{{Hahn} \& {Abel}(2011)}]{hahn11}
{Hahn}, O., \& {Abel}, T. 2011, \mnras, 415, 2101,
  \dodoi{10.1111/j.1365-2966.2011.18820.x}

\bibitem[{{Herrington} {et~al.}(2023){Herrington}, {Whalen}, \&
  {Woods}}]{herr23a}
{Herrington}, N.~P., {Whalen}, D.~J., \& {Woods}, T.~E. 2023, \mnras, 521, 463,
  \dodoi{10.1093/mnras/stad572}

\bibitem[{{Hosokawa} {et~al.}(2013){Hosokawa}, {Yorke}, {Inayoshi}, {Omukai},
  \& {Yoshida}}]{hos13}
{Hosokawa}, T., {Yorke}, H.~W., {Inayoshi}, K., {Omukai}, K., \& {Yoshida}, N.
  2013, \apj, 778, 178, \dodoi{10.1088/0004-637X/778/2/178}

\bibitem[{{Hviding} {et~al.}(2025){Hviding}, {de Graaff}, {Miller}, {Setton},
  {Greene}, {Labb{\'e}}, {Brammer}, {Bezanson}, {Boogaard}, {Cleri}, {Leja},
  {Maseda}, {McConachie}, {Matthee}, {Naidu}, {Oesch}, {Wang}, {Whitaker}, \&
  {Williams}}]{hvid25}
{Hviding}, R.~E., {de Graaff}, A., {Miller}, T.~B., {et~al.} 2025, \aap, 702,
  A57, \dodoi{10.1051/0004-6361/202555816}

\bibitem[{{Inayoshi} \& {Maiolino}(2025)}]{Inayoshi2025}
{Inayoshi}, K., \& {Maiolino}, R. 2025, \apjl, 980, L27,
  \dodoi{10.3847/2041-8213/adaebd}

\bibitem[{{Jeon} {et~al.}(2025){Jeon}, {Bromm}, {Liu}, \&
  {Finkelstein}}]{jeon25}
{Jeon}, J., {Bromm}, V., {Liu}, B., \& {Finkelstein}, S.~L. 2025, \apj, 979,
  127, \dodoi{10.3847/1538-4357/ad9f3a}

\bibitem[{{Jeon} {et~al.}(2026){Jeon}, {Liu}, {Bromm}, {Fujimoto}, {Taylor},
  {Kokorev}, {Larson}, {Chisholm}, {Finkelstein}, \& {Kocevski}}]{jeon26}
{Jeon}, J., {Liu}, B., {Bromm}, V., {et~al.} 2026, \apj, 998, 148,
  \dodoi{10.3847/1538-4357/ae3725}

\bibitem[{{Kim} {et~al.}(2011){Kim}, {Wise}, {Alvarez}, \& {Abel}}]{kim11}
{Kim}, J.-h., {Wise}, J.~H., {Alvarez}, M.~A., \& {Abel}, T. 2011, \apj, 738,
  54, \dodoi{10.1088/0004-637X/738/1/54}

\bibitem[{{Kocevski} {et~al.}(2025){Kocevski}, {Finkelstein}, {Barro},
  {Taylor}, {Calabr{\`o}}, {Laloux}, {Buchner}, {Trump}, {Leung}, {Yang},
  {Dickinson}, {P{\'e}rez-Gonz{\'a}lez}, {Pacucci}, {Inayoshi}, {Somerville},
  {McGrath}, {Akins}, {Bagley}, {Bowler}, {Bisigello}, {Carnall}, {Casey},
  {Cheng}, {Cleri}, {Costantin}, {Cullen}, {Davis}, {Donnan}, {Dunlop},
  {Ellis}, {Ferguson}, {Fujimoto}, {Fontana}, {Giavalisco}, {Grazian},
  {Grogin}, {Hathi}, {Hirschmann}, {Huertas-Company}, {Holwerda},
  {Illingworth}, {Juneau}, {Kartaltepe}, {Koekemoer}, {Li}, {Lucas}, {Magee},
  {Mason}, {McLeod}, {McLure}, {Napolitano}, {Papovich}, {Pirzkal},
  {Rodighiero}, {Santini}, {Wilkins}, \& {Yung}}]{lrd4}
{Kocevski}, D.~D., {Finkelstein}, S.~L., {Barro}, G., {et~al.} 2025, \apj, 986,
  126, \dodoi{10.3847/1538-4357/adbc7d}

\bibitem[{{Kov{\'a}cs} {et~al.}(2024){Kov{\'a}cs}, {Bogd{\'a}n}, {Natarajan},
  {Werner}, {Azadi}, {Volonteri}, {Tremblay}, {Chadayammuri}, {Forman},
  {Jones}, \& {Kraft}}]{Kov24}
{Kov{\'a}cs}, O.~E., {Bogd{\'a}n}, {\'A}., {Natarajan}, P., {et~al.} 2024,
  \apjl, 965, L21, \dodoi{10.3847/2041-8213/ad391f}

\bibitem[{{Latif} {et~al.}(2024){Latif}, {Aftab}, \& {Whalen}}]{latif24b}
{Latif}, M.~A., {Aftab}, A., \& {Whalen}, D.~J. 2024, \aj, 167, 251,
  \dodoi{10.3847/1538-3881/ad3b97}

\bibitem[{{Latif} {et~al.}(2025){Latif}, {Aftab}, {Whalen}, \&
  {Mezcua}}]{latif25a}
{Latif}, M.~A., {Aftab}, A., {Whalen}, D.~J., \& {Mezcua}, M. 2025, \aap, 694,
  L14, \dodoi{10.1051/0004-6361/202453194}

\bibitem[{{Latif} \& {Khochfar}(2020)}]{latif20b}
{Latif}, M.~A., \& {Khochfar}, S. 2020, \mnras, 497, 3761,
  \dodoi{10.1093/mnras/staa2218}

\bibitem[{{Latif} {et~al.}(2021){Latif}, {Khochfar}, {Schleicher}, \&
  {Whalen}}]{latif21a}
{Latif}, M.~A., {Khochfar}, S., {Schleicher}, D., \& {Whalen}, D.~J. 2021,
  \mnras, 508, 1756, \dodoi{10.1093/mnras/stab2708}

\bibitem[{{Latif} \& {Whalen}(2025)}]{L25}
{Latif}, M.~A., \& {Whalen}, D.~J. 2025, \apjl, 990, L58,
  \dodoi{10.3847/2041-8213/adfec6}

\bibitem[{{Latif} {et~al.}(2026){Latif}, {Whalen}, {Khochfar}, \&
  {Cullen}}]{latif26a}
{Latif}, M.~A., {Whalen}, D.~J., {Khochfar}, S., \& {Cullen}, F. 2026, \apjl,
  1003, L40, \dodoi{10.3847/2041-8213/ae67f0}

\bibitem[{{Latif} {et~al.}(2022){Latif}, {Whalen}, {Khochfar}, {Herrington}, \&
  {Woods}}]{latif22b}
{Latif}, M.~A., {Whalen}, D.~J., {Khochfar}, S., {Herrington}, N.~P., \&
  {Woods}, T.~E. 2022, \nat, 607, 48, \dodoi{10.1038/s41586-022-04813-y}

\bibitem[{{Latif} {et~al.}(2011){Latif}, {Zaroubi}, \& {Spaans}}]{latif11}
{Latif}, M.~A., {Zaroubi}, S., \& {Spaans}, M. 2011, \mnras, 411, 1659,
  \dodoi{10.1111/j.1365-2966.2010.17796.x}

\bibitem[{{Liu} {et~al.}(2025){Liu}, {Jiang}, {Quataert}, {Greene}, \&
  {Ma}}]{Liu2025}
{Liu}, H., {Jiang}, Y.-F., {Quataert}, E., {Greene}, J.~E., \& {Ma}, Y. 2025,
  \apj, 994, 113, \dodoi{10.3847/1538-4357/ae0c19}

\bibitem[{{Maiolino} {et~al.}(2025){Maiolino}, {Risaliti}, {Signorini},
  {Trefoloni}, {Juod{\v{z}}balis}, {Scholtz}, {{\"U}bler}, {D'Eugenio},
  {Carniani}, {Fabian}, {Ji}, {Mazzolari}, {Bertola}, {Brusa}, {Bunker},
  {Charlot}, {Comastri}, {Cresci}, {DeCoursey}, {Egami}, {Fiore}, {Gilli},
  {Perna}, {Tacchella}, \& {Venturi}}]{Maiolino2025}
{Maiolino}, R., {Risaliti}, G., {Signorini}, M., {et~al.} 2025, \mnras, 538,
  1921, \dodoi{10.1093/mnras/staf359}

\bibitem[{{Matthee} {et~al.}(2024){Matthee}, {Naidu}, {Brammer}, {Chisholm},
  {Eilers}, {Goulding}, {Greene}, {Kashino}, {Labbe}, {Lilly}, {Mackenzie},
  {Oesch}, {Weibel}, {Wuyts}, {Xiao}, {Bordoloi}, {Bouwens}, {van Dokkum},
  {Illingworth}, {Kramarenko}, {Maseda}, {Mason}, {Meyer}, {Nelson}, {Reddy},
  {Shivaei}, {Simcoe}, \& {Yue}}]{lrd1}
{Matthee}, J., {Naidu}, R.~P., {Brammer}, G., {et~al.} 2024, \apj, 963, 129,
  \dodoi{10.3847/1538-4357/ad2345}

\bibitem[{{Matthee} {et~al.}(2026){Matthee}, {Torralba}, {Pezzulli}, {Naidu},
  {Chisholm}, {Mascia}, {Greene}, {Ishikawa}, {Gronke}, {Wuyts}, {Bordoloi},
  {Brammer}, {Chang}, {Eilers}, {de Graaff}, {Hviding}, {Iani}, {Illingworth},
  {Kashino}, {Labbe}, {Ma}, {Maseda}, {Meyer}, {Nelson}, {Oesch}, \&
  {Xiao}}]{Matthee2026}
{Matthee}, J., {Torralba}, A., {Pezzulli}, G., {et~al.} 2026, arXiv e-prints,
  arXiv:2603.17667, \dodoi{10.48550/arXiv.2603.17667}

\bibitem[{{Naidu} {et~al.}(2026){Naidu}, {Matthee}, {Katz}, {de Graaff},
  {Oesch}, {Smith}, {Greene}, {Brammer}, {Weibel}, {Hviding}, {Chisholm},
  {Labb{\'e}}, {Simcoe}, {Witten}, {Sun}, {Atek}, {Baggen}, {Belli},
  {Bezanson}, {Boogaard}, {Bose}, {Bouwens}, {Covelo-Paz}, {Dayal}, {Fudamoto},
  {Furtak}, {Giovinazzo}, {Goulding}, {Gronke}, {Heintz}, {Hirschmann},
  {Illingworth}, {Inoue}, {Johnson}, {Leja}, {Leonova}, {McConachie}, {Maseda},
  {Natarajan}, {Nelson}, {Setton}, {Shivaei}, {Sobral}, {Stefanon},
  {Tacchella}, {Toft}, {Torralba}, {van Dokkum}, {van der Wel}, {Volonteri},
  {Walter}, {Wang}, {Watson}, \& {Whitaker}}]{nai26}
{Naidu}, R.~P., {Matthee}, J., {Katz}, H., {et~al.} 2026, \nat, 656, 329,
  \dodoi{10.1038/s41586-026-10846-4}

\bibitem[{{Nandal} {et~al.}(2026){Nandal}, {Chilingarian}, {Nagele},
  {Chisholm}, {Bauer}, \& {Loeb}}]{nan26b}
{Nandal}, D., {Chilingarian}, I.~V., {Nagele}, C., {et~al.} 2026, \apjl, 1006,
  L21, \dodoi{10.3847/2041-8213/ae82f3}

\bibitem[{{Nandal} \& {Loeb}(2026)}]{nan26a}
{Nandal}, D., \& {Loeb}, A. 2026, \apj, 998, 124,
  \dodoi{10.3847/1538-4357/ae32f3}

\bibitem[{{Nandal} {et~al.}(2025){Nandal}, {Whalen}, {Latif}, \&
  {Heger}}]{nan25a}
{Nandal}, D., {Whalen}, D.~J., {Latif}, M.~A., \& {Heger}, A. 2025, \apjl, 994,
  L11, \dodoi{10.3847/2041-8213/ae1a63}

\bibitem[{{Nandal} {et~al.}(2024){Nandal}, {Zwick}, {Whalen}, {Mayer},
  {Ekstr{\"o}m}, \& {Meynet}}]{nan24d}
{Nandal}, D., {Zwick}, L., {Whalen}, D.~J., {et~al.} 2024, \aap, 689, A351,
  \dodoi{10.1051/0004-6361/202449562}

\bibitem[{{Nanni} {et~al.}(2017){Nanni}, {Vignali}, {Gilli}, {Moretti}, \&
  {Brandt}}]{nan17}
{Nanni}, R., {Vignali}, C., {Gilli}, R., {Moretti}, A., \& {Brandt}, W.~N.
  2017, \aap, 603, A128, \dodoi{10.1051/0004-6361/201730484}

\bibitem[{{Pacucci} {et~al.}(2026){Pacucci}, {Ferrara}, \& {Kocevski}}]{pac26}
{Pacucci}, F., {Ferrara}, A., \& {Kocevski}, D.~D. 2026, arXiv e-prints,
  arXiv:2601.14368, \dodoi{10.48550/arXiv.2601.14368}

\bibitem[{{Patrick} {et~al.}(2023){Patrick}, {Whalen}, {Latif}, \&
  {Elford}}]{pat23a}
{Patrick}, S.~J., {Whalen}, D.~J., {Latif}, M.~A., \& {Elford}, J.~S. 2023,
  \mnras, 522, 3795, \dodoi{10.1093/mnras/stad1179}

\bibitem[{{Planck Collaboration} {et~al.}(2016){Planck Collaboration}, {Ade},
  {Aghanim}, {Arnaud}, {Ashdown}, {Aumont}, {Baccigalupi}, {Banday},
  {Barreiro}, {Bartlett}, \& et~al.}]{planck2}
{Planck Collaboration}, {Ade}, P.~A.~R., {Aghanim}, N., {et~al.} 2016, \aap,
  594, A13, \dodoi{10.1051/0004-6361/201525830}

\bibitem[{{Regan} {et~al.}(2017){Regan}, {Visbal}, {Wise}, {Haiman},
  {Johansson}, \& {Bryan}}]{regan17a}
{Regan}, J.~A., {Visbal}, E., {Wise}, J.~H., {et~al.} 2017, Nature Astronomy,
  1, 0075, \dodoi{10.1038/s41550-017-0075}

\bibitem[{{Rusakov} {et~al.}(2026{\natexlab{a}}){Rusakov}, {Watson},
  {Nikopoulos}, {Brammer}, {Gottumukkala}, {Harvey}, {Heintz}, {Damgaard},
  {Sim}, {Sneppen}, {Vijayan}, {Adams}, {Austin}, {Conselice}, {Goolsby},
  {Toft}, \& {Witstok}}]{Rusakov2026}
{Rusakov}, V., {Watson}, D., {Nikopoulos}, G.~P., {et~al.} 2026{\natexlab{a}},
  \nat, 649, 574, \dodoi{10.1038/s41586-025-09900-4}

\bibitem[{{Rusakov} {et~al.}(2026{\natexlab{b}}){Rusakov}, {Watson},
  {Nikopoulos}, {Brammer}, {Gottumukkala}, {Harvey}, {Heintz}, {Damgaard},
  {Sim}, {Sneppen}, {Vijayan}, {Adams}, {Austin}, {Conselice}, {Goolsby},
  {Toft}, \& {Witstok}}]{rus26}
---. 2026{\natexlab{b}}, \nat, 649, 574, \dodoi{10.1038/s41586-025-09900-4}

\bibitem[{{Sazonov} {et~al.}(2004){Sazonov}, {Ostriker}, \& {Sunyaev}}]{sos04}
{Sazonov}, S.~Y., {Ostriker}, J.~P., \& {Sunyaev}, R.~A. 2004, \mnras, 347,
  144, \dodoi{10.1111/j.1365-2966.2004.07184.x}

\bibitem[{{Setton} {et~al.}(2024){Setton}, {Khullar}, {Miller}, {Bezanson},
  {Greene}, {Suess}, {Whitaker}, {Antwi-Danso}, {Atek}, {Brammer}, {Cutler},
  {Dayal}, {Feldmann}, {Fujimoto}, {Furtak}, {Glazebrook}, {Goulding},
  {Kokorev}, {Labbe}, {Leja}, {Ma}, {Marchesini}, {Nanayakkara}, {Pan},
  {Price}, {Siegel}, {Shipley}, {Weaver}, {van Dokkum}, {Wang}, \&
  {Williams}}]{Setton2024}
{Setton}, D.~J., {Khullar}, G., {Miller}, T.~B., {et~al.} 2024, \apj, 974, 145,
  \dodoi{10.3847/1538-4357/ad6a18}

\bibitem[{{Setton} {et~al.}(2025){Setton}, {Greene}, {de Graaff}, {Ma}, {Leja},
  {Matthee}, {Bezanson}, {Boogaard}, {Cleri}, {Katz}, {Labbe}, {Maseda},
  {McConachie}, {Miller}, {Price}, {Suess}, {van Dokkum}, {Wang}, {Weibel},
  {Whitaker}, \& {Williams}}]{Setton2025}
{Setton}, D.~J., {Greene}, J.~E., {de Graaff}, A., {et~al.} 2025, \apj, 995,
  118, \dodoi{10.3847/1538-4357/ae1500}

\bibitem[{{Shull} \& {van Steenberg}(1985)}]{Shull85}
{Shull}, J.~M., \& {van Steenberg}, M.~E. 1985, \apj, 298, 268,
  \dodoi{10.1086/163605}

\bibitem[{{Smidt} {et~al.}(2018){Smidt}, {Whalen}, {Johnson}, {Surace}, \&
  {Li}}]{smidt18}
{Smidt}, J., {Whalen}, D.~J., {Johnson}, J.~L., {Surace}, M., \& {Li}, H. 2018,
  \apj, 865, 126, \dodoi{10.3847/1538-4357/aad7b8}

\bibitem[{{Smith} {et~al.}(2017){Smith}, {Becerra}, {Bromm}, \&
  {Hernquist}}]{aaron17}
{Smith}, A., {Becerra}, F., {Bromm}, V., \& {Hernquist}, L. 2017, \mnras, 472,
  205, \dodoi{10.1093/mnras/stx1993}

\bibitem[{{Sneppen} {et~al.}(2026){Sneppen}, {Watson}, {Matthews},
  {Nikopoulos}, {Allen}, {Brammer}, {Damgaard}, {Heintz}, {Knigge}, {Long},
  {Rusakov}, {Sim}, \& {Witstok}}]{snep26}
{Sneppen}, A., {Watson}, D., {Matthews}, J.~H., {et~al.} 2026, arXiv e-prints,
  arXiv:2601.18864, \dodoi{10.48550/arXiv.2601.18864}

\bibitem[{{Sun} {et~al.}(2026){Sun}, {Naidu}, {Matthee}, {de Graaff},
  {Chisholm}, {Greene}, {Oesch}, {Torralba}, {Hviding}, {Brammer}, {Simcoe},
  {Bose}, {Bouwens}, {Dayal}, {Eilers}, {Fei}, {Furtak}, {Gottumukkala},
  {Goulding}, {Heintz}, {Hirschmann}, {Kokorev}, {Leja}, {Liu}, {Natarajan},
  {Santarelli}, {Setton}, {Smith}, {Tacchella}, {Volonteri}, {Walter},
  {Weibel}, \& {Williams}}]{sun26}
{Sun}, W.~Q., {Naidu}, R.~P., {Matthee}, J., {et~al.} 2026, The Open Journal of
  Astrophysics, 9, 62505, \dodoi{10.33232/001c.162505}

\bibitem[{{Surace} {et~al.}(2019){Surace}, {Zackrisson}, {Whalen}, {Hartwig},
  {Glover}, {Woods}, {Heger}, \& {Glover}}]{sur19a}
{Surace}, M., {Zackrisson}, E., {Whalen}, D.~J., {et~al.} 2019, \mnras, 488,
  3995, \dodoi{10.1093/mnras/stz1956}

\bibitem[{{Surace} {et~al.}(2018){Surace}, {Whalen}, {Hartwig}, {Zackrisson},
  {Glover}, {Patrick}, {Woods}, {Heger}, \& {Haemmerl{\'e}}}]{sur18a}
{Surace}, M., {Whalen}, D.~J., {Hartwig}, T., {et~al.} 2018, \apjl, 869, L39,
  \dodoi{10.3847/2041-8213/aaf80d}

\bibitem[{{Taylor} {et~al.}(2025){Taylor}, {Kokorev}, {Kocevski}, {Akins},
  {Cullen}, {Dickinson}, {Finkelstein}, {Arrabal Haro}, {Bromm}, {Giavalisco},
  {Inayoshi}, {Juneau}, {Leung}, {P{\'e}rez-Gonz{\'a}lez}, {Somerville},
  {Trump}, {Amor{\'\i}n}, {Barro}, {Burgarella}, {Brooks}, {Carnall}, {Casey},
  {Cheng}, {Chisholm}, {Chworowsky}, {Davis}, {Donnan}, {Dunlop}, {Ellis},
  {Fern{\'a}ndez}, {Fujimoto}, {Grogin}, {Gupta}, {Hathi}, {Jung},
  {Hirschmann}, {Kartaltepe}, {Koekemoer}, {Larson}, {Leung}, {Llerena},
  {Lucas}, {McLeod}, {McLure}, {Napolitano}, {Papovich}, {Stanton}, {Tripodi},
  {Wang}, {Wilkins}, {Yung}, \& {Zavala}}]{tyl25}
{Taylor}, A.~J., {Kokorev}, V., {Kocevski}, D.~D., {et~al.} 2025, \apjl, 989,
  L7, \dodoi{10.3847/2041-8213/ade789}

\bibitem[{{Toro} {et~al.}(1994){Toro}, {Spruce}, \& {Speares}}]{toro94}
{Toro}, E.~F., {Spruce}, M., \& {Speares}, W. 1994, Shock Waves, 4, 25,
  \dodoi{10.1007/BF01414629}

\bibitem[{{Torralba} {et~al.}(2026){Torralba}, {Matthee}, {Pezzulli}, {Naidu},
  {Ishikawa}, {Brammer}, {Chang}, {Chisholm}, {de Graaff}, {D'Eugenio}, {Di
  Cesare}, {Eilers}, {Greene}, {Gronke}, {Iani}, {Kokorev}, {Kotiwale},
  {Kramarenko}, {Ma}, {Mascia}, {Navarrete}, {Nelson}, {Oesch}, {Simcoe}, \&
  {Wuyts}}]{Torralba2025}
{Torralba}, A., {Matthee}, J., {Pezzulli}, G., {et~al.} 2026, \aap, 707, A75,
  \dodoi{10.1051/0004-6361/202557537}

\bibitem[{{Whalen} \& {Fryer}(2012)}]{wf12}
{Whalen}, D.~J., \& {Fryer}, C.~L. 2012, \apjl, 756, L19,
  \dodoi{10.1088/2041-8205/756/1/L19}

\bibitem[{{Whalen} {et~al.}(2021){Whalen}, {Mezcua}, {Patrick}, {Meiksin}, \&
  {Latif}}]{wet21a}
{Whalen}, D.~J., {Mezcua}, M., {Patrick}, S.~J., {Meiksin}, A., \& {Latif},
  M.~A. 2021, \apjl, 922, L39, \dodoi{10.3847/2041-8213/ac35e6}

\bibitem[{{Whalen} {et~al.}(2020){Whalen}, {Surace}, {Bernhardt}, {Zackrisson},
  {Pacucci}, {Ziegler}, \& {Hirschmann}}]{wet20b}
{Whalen}, D.~J., {Surace}, M., {Bernhardt}, C., {et~al.} 2020, \apjl, 897, L16,
  \dodoi{10.3847/2041-8213/ab9d29}

\bibitem[{{Wise} \& {Abel}(2011)}]{moray}
{Wise}, J.~H., \& {Abel}, T. 2011, \mnras, 414, 3458,
  \dodoi{10.1111/j.1365-2966.2011.18646.x}

\bibitem[{{Woods} {et~al.}(2017){Woods}, {Heger}, {Whalen}, {Haemmerl{\'e}}, \&
  {Klessen}}]{tyr17}
{Woods}, T.~E., {Heger}, A., {Whalen}, D.~J., {Haemmerl{\'e}}, L., \&
  {Klessen}, R.~S. 2017, \apjl, 842, L6, \dodoi{10.3847/2041-8213/aa7412}

\bibitem[{{Yue} {et~al.}(2013){Yue}, {Ferrara}, {Salvaterra}, {Xu}, \&
  {Chen}}]{yue13}
{Yue}, B., {Ferrara}, A., {Salvaterra}, R., {Xu}, Y., \& {Chen}, X. 2013,
  \mnras, 433, 1556, \dodoi{10.1093/mnras/stt826}

\end{thebibliography}
\bibliographystyle{aasjournal}

\end{document}